\documentclass[aps, prl,twocolumn,showpacs,superscriptaddress]{revtex4-2}

\usepackage{graphicx}
\usepackage{mathrsfs}
\usepackage{dcolumn}
\usepackage{bm}
\usepackage{amsmath}
\usepackage{amsfonts}
\usepackage{color}
\usepackage[colorlinks=true,linkcolor=blue,
            pagecolor=blue,filecolor=blue,
            menucolor=blue,urlcolor=blue,
            citecolor=blue,anchorcolor=blue]{hyperref}
\usepackage{lineno}
\usepackage{tabularx}
\usepackage{epstopdf}
\usepackage{booktabs}
\usepackage{float}
\usepackage{multirow}
\usepackage{array}
\usepackage{makecell}
\usepackage{gensymb}

\makeatletter

\newcommand{\Rmnum}[1]{\expandafter\@slowromancap\romannumeral #1@}
\makeatother

\begin{document}

    \title{ Structural instability and charge modulations in the Kagome superconductor $A$V$_3$Sb$_5$}

    \author{Zijin Ye}
    \affiliation{Wuhan National High Magnetic Field Center and School of Physics, Huazhong University of Science and Technology, Wuhan 430074, China}
    \author{Aiyun Luo}
    \affiliation{Wuhan National High Magnetic Field Center and School of Physics, Huazhong University of Science and Technology, Wuhan 430074, China}
    \author{Jia-Xin Yin}
    \affiliation{Laboratory for Topological Quantum Matter and Advanced Spectroscopy (B7),Department of Physics, Princeton University, Princeton, New Jersey 08544, USA}
    \author{M Zahid Hasan}
    \affiliation{Laboratory for Topological Quantum Matter and Advanced Spectroscopy (B7),Department of Physics, Princeton University, Princeton, New Jersey 08544, USA}
    \author{Gang Xu}
    \email{gangxu@hust.edu.cn}
    \affiliation{Wuhan National High Magnetic Field Center and School of Physics, Huazhong University of Science and Technology, Wuhan 430074, China}

    \date{\today}

\begin{abstract}
    Recently, both charge density wave (CDW) and superconductivity have been observed in Kagome compounds $A$V$_3$Sb$_5$. However, the nature of CDW that results in many novel charge modulations is still under hot debate.
    By means of the first-principles calculations, we discover two kinds of CDW states, the trimerized and hexamerized 2$\times$2 phase and dimerized 4$\times$1 phase existing in $A$V$_3$Sb$_5$.
    Our phonon excitation spectrum and electronic Lindhard function calculations reveal that the most intensive structural instability in $A$V$_3$Sb$_5$ originates from a combined in-plane vibration mode of V atoms through the electron-phonon coupling, rather than the Fermi surface nesting effect.
    Crucially, a metastable 4$\times$1 phase with V-V dimer pattern and twofold symmetric bowtie shaped charge modulation is revealed in CsV$_3$Sb$_5$, implying that both dimerization and trimerization exist in the V Kagome layers.
    These results provide essential understanding of CDW instability and new thoughts for the novel charge modulation patterns.

\end{abstract}

\maketitle

    Kagome materials are the unique platform to study the topological physics~\cite{PhysRevLett.115.186802}, flat band~\cite{RN36}, geometrical spin frustration~\cite{norman2016colloquium} and their interactions.
    At the beginning, large band gap Kagome magnets are widely studied as the promising quantum spin liquid system~\cite{Olariu2008,RN29,yan2011spin,han2012fractionalized,shen2016evidence}.
    Encouraged by the quantum anomalous Hall (QAH) effect proposed in the ferromagnetic Kagome material Cs$_2$LiMn$_3$F$_{12}$~\cite{PhysRevLett.115.186802}, the topological physics~\cite{ye2018massive,yin2020quantum,tsai2020electrical}, electronic correlated flat band~\cite{RN36,kang2020dirac} and quantum transport induced by the nontrivial Berry curvature~\cite{nakatsuji2015large,liu2018giant} in the Kagome metals have attracted increasing interest.
    Additionally, the interplay between electronic correlation, magnetic frustration and topology usually gives rise to a variety of intriguing quantum phenomena~\cite{yin2018giant}, such as fractional QAH effect~\cite{PhysRevLett.106.236802}, topological phase transition~\cite{He2021}, spin or charge density waves (SDWs or CDWs)~\cite{PhysRevB.85.144402,PhysRevB.87.115135} and superconductivity~\cite{PhysRevB.86.121105,PhysRevLett.110.126405,Mielke2021}.
    Recently, a new family of quasi-two-dimensional (quasi-2D) Kagome metal $A$V$_3$Sb$_5$ ($A$ = K, Rb, Cs) has been synthesized~\cite{PhysRevMaterials.3.094407}, and reported to host nontrivial topological bands~\cite{Yangeabb6003}, unconventional superconductivity~\cite{PhysRevLett.125.247002,wang2020proximityinduced}, van Hove singularities (VHS) and CDW~\cite{RN40,Yin_2021,PhysRevLett.127.046401,FENG20211384,PhysRevB.103.L241117}, which provide a natural platform to study the interplay between these quantum states.

\begin{figure*}
 	\centering
 	\includegraphics[width=1\textwidth]{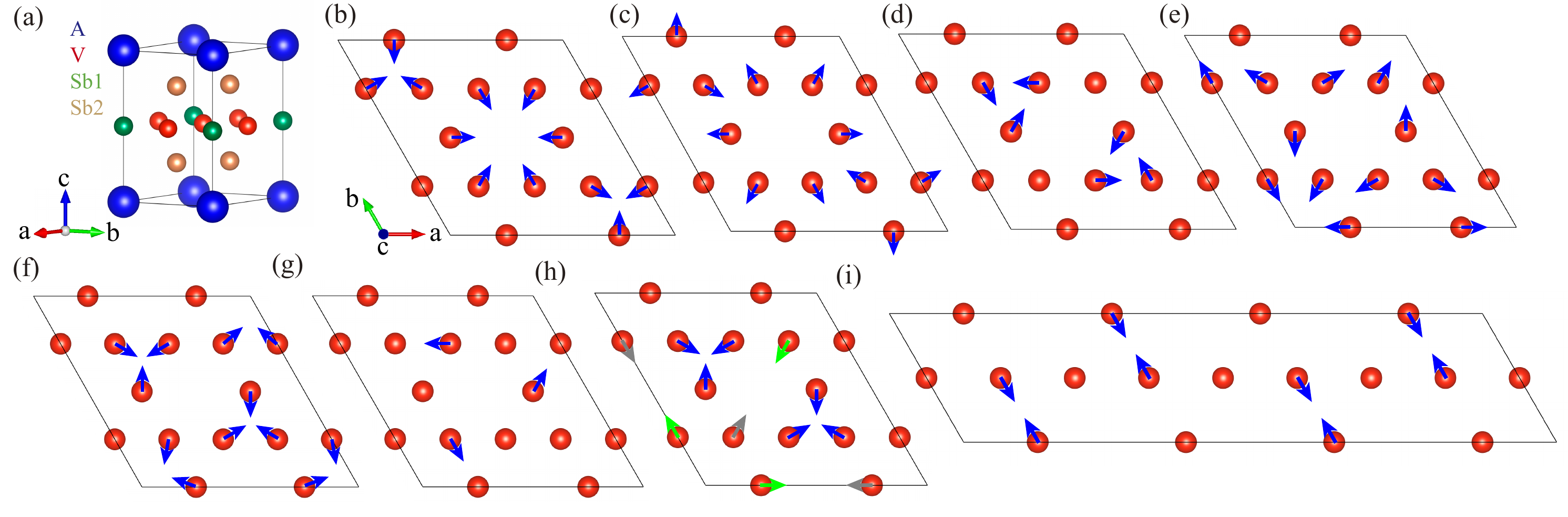}
 	\caption{
 		(a) The crystal structure of pristine $A$V$_3$Sb$_5$.
 		(b)-(i) The top view along the $c$-axis of V Kagome layer in the 2$\times$2 supercells satisfying $D_{6h}$((b) for T\&H, (c) for DS), $C_{6h}$ (d), $D_{3d}$ (e), $D_{3h}$ (f), $C_{3h}$ (g), $C_{3v}$ (h) and 4$\times$1 Dimer phase (i), respectively.
         The blue arrows represent pure in-plane movements, and the green (gray) arrows represent in-plane movements with a slight upward (downward) shift along $c$ direction.
 	}
 	\label{fig1}
 \end{figure*}

    As shown in Fig.~1(a), the pristine $A$V$_3$Sb$_5$ compounds share same layered structure satisfying space group $P6/mmm$ (No. 191),
    where $A$ atoms at $1a$ (0, 0, 0) site form the triangle lattice, V atoms at $3g$ ($\frac{1}{2}$, 0, $\frac{1}{2}$) site form a Kagome layer,
    Sb1 atoms at $1b$ (0, 0, $\frac{1}{2}$) occupy the center of V hexagon and Sb2 atoms at $4h$ ($\frac{1}{3}$, $\frac{2}{3}$, $z$) form two honeycomb layers.
    These materials undergo a CDW transition as cooling down to about 80 - 100 K~\cite{PhysRevLett.125.247002,RN40,Yin_2021}, and enter to the superconducting phase with $T_c$ = 0.9 - 2.5 K~\cite{PhysRevLett.125.247002,wang2020proximityinduced,PhysRevMaterials.5.034801}.
    For the CDW phase, a 2$\times$2 or 2$\times$2$\times$2 superlattice with trimerization and hexamerization of V atoms (referred to as T\&H phase) is discovered in KV$_3$Sb$_5$~\cite{RN40} and CsV$_3$Sb$_5$~\cite{PhysRevLett.127.046401,PhysRevX.11.031026}.
    Very recently, an additional 4$\times$1 CDW phase is further observed by scanning tunneling microscopy (STM) experiments in CsV$_3$Sb$_5$ when the samples are cooled down to 50 K~\cite{PhysRevB.104.075148,RN42,RN41}.
    Some theoretical and experimental works suggested that the CDW might be originated from the electron correlation~\cite{PhysRevB.87.115135,PhysRevB.77.165135,PhysRevLett.25.919,PhysRevLett.107.107403,Zhu2367,PhysRevB.103.115135,PhysRevB.40.7391,MACHIDA1989192,PhysRevB.39.9749,PhysRevLett.51.138}.
    Some X-ray scattering and angle-resolved photoemission spectroscopy (ARPES) experiments suggest that it is purely induced by the Fermi surface (FS) nesting of the VHS at $M$ points~\cite{PhysRevX.11.031050}.
    Besides, recent ARPES and neutron scattering experiments propose that the 2$\times$2 CDW phase in $A$V$_3$Sb$_5$ is mainly driven by the electron-phonon coupling (EPC)~\cite{luo2021electronic,cho2021emergence,wang2021distinctive,nakayama2021multiple,xie2021electronphonon}.
    For the 4$\times$1 CDW, Zhao \emph{et al.} propose that it may be associated with the bulk electronic nematicity~\cite{RN41,RN42},
    while Li \emph{et al.} attribute it to the surface instability and electron correlation ~\cite{li2021spatial}.
    Therefore, the nature of CDW and its relation to the electronic structures, EPC, electron correlation and superconductivity remain under hot debate in $A$V$_3$Sb$_5$.

    In this paper, we construct a variety of supercells to study the structural instability and charge modulation in $A$V$_3$Sb$_5$ based on the first-principles calculations.
    Our calculations reveal that the most stable structure in $A$V$_3$Sb$_5$ is the 2$\times$2 T\&H phase, and the reconstruction along $c$-direction is negligible.
    Further phonon spectrum and Lindhard function calculations demonstrate that such CDW instability is mainly driven by a combined in-plane vibration mode of V atoms, rather than the FS nesting effect.
    Such vibration gives rise to a negative phonon mode at $M$ point, which can be hardened after the trimerization and hexamerization of V atoms in the T\&H phase.
    More importantly, a metastable 4$\times$1 phase with V-V dimer pattern is also discovered in CsV$_3$Sb$_5$, leading to a twofold symmetric bowtie shaped charge modulation that has never been observed previously.
    The bowtie shaped charge modulation may be responsible to the twofold resistivity anisotropy in CsV$_3$Sb$_5$~\cite{RN42}.
    Our results demonstrate that the competition between dimerization and trimerization may exist in CsV$_3$Sb$_5$, which shield lights on the essential understanding of CDW instability and other novel phenomena.

    Our first-principles calculations are carried out by the Vienna ab initio simulation package (VASP)~\cite{PhysRevB.54.11169,PhysRevB.48.13115,PhysRevB.59.1758}.
    The generalized gradient approximation (GGA) of the Perdew-Burke-Ernzerhof (PBE) type is adopted for the exchange-correlation potential~\cite{PhysRevLett.77.3865}.
    The cut-off energy for the wave function expansion is set to 500 eV and 16$\times$16$\times$9 k-mesh is used to sample the first Brillouin zone (BZ).
    The experimental lattice parameters $a = 5.482 \AA$, $c = 8.958 \AA$ for KV$_3$Sb$_5$ and $a = 5.44 \AA$, $c = 9.33 \AA$ for CsV$_3$Sb$_5$ are used~\cite{PhysRevLett.127.046401},
    and all structures are optimized until the force on each atom is less than 0.01 eV/\AA.
    The unfolded band structures of CsV$_3$Sb$_5$ are obtained from the Wannier functions by using WANNIER90~\cite{MOSTOFI20142309} and WannierTools package~\cite{WU2018405}.

\begin{table}
	\caption{
		The energy of KV$_3$Sb$_5$ in different strucutures.
	}
	\begin{tabularx}{22em}%
		{>{\centering\arraybackslash}X>{\centering\arraybackslash}X}
		\toprule[1pt]
		Structure & Energy (eV/f.u.) \\ \hline
		pristine  & -50.511          \\ 
		T\&H      & -50.519   		 \\ 
		DS        & -50.514   	 	 \\ 
		$C_{6h}$  & -50.518   		 \\ 
		$D_{3d}$  & -50.514   		 \\ 
		$D_{3h}$  & -50.518   		 \\ 
		$C_{3h}$  & -50.514   		 \\ 
		$C_{3v}$  & -50.514   		 \\ \bottomrule[1pt]
	\end{tabularx}
\end{table}

\begin{table}
	\caption{
		The energy of CsV$_3$Sb$_5$ in different structures.
	}
	\begin{tabularx}{20em}{cp{0.2cm}cp{0.2cm}c}
		\toprule[1pt]
		Structure                        & & \multicolumn{3}{c}{Energy (eV/f.u.)}          \\ \hline
		pristine                         & & \multicolumn{3}{c}{-50.571}                   \\ \hline
				                         & & (2 $\times$ 2) & & (2 $\times$ 2 $\times$ 2 ) \\ 
		T\&H                             & & -50.586        & & -50.586         		   \\ 
		$C_{6h}$                         & & -50.585        & & -50.585 	               \\ 
		$D_{3d}$                         & & -50.575        & & -50.575     		       \\ 
		$C_{3v}$                         & & -50.575        & & -50.575         		   \\ \hline
		4$\times$1 Dimer            & & \multicolumn{3}{c}{-50.574} 		           \\
		4$\times$1$\times$2 Dimer & & \multicolumn{3}{c}{-50.574} 		           \\ \bottomrule[1pt]
	\end{tabularx}
\end{table}

    Considering that many experimental results indicate the time reversal symmetry breaking in $A$V$_3$Sb$_5$~\cite{RN40,SongDW2021,PhysRevB.104.L041103,MielkeC2021,Yangeabb6003,YuL2021,PhysRevB.104.035131}, we study their magnetic instability by calculating many magnetic configurations.
    However, their energies always return to that of non-magnetic (NM) state [SM],
    indicating that any magnetism has not been found in the Density Functional Theory level.
    These results are consistent with recent experiments~\cite{PhysRevMaterials.3.094407,10.1088/1361-648X/abe8f9},
    implying that time reversal symmetry breaking effect may come from CDW transition or electron correlation.

\begin{figure}
	\includegraphics[width=0.48\textwidth]{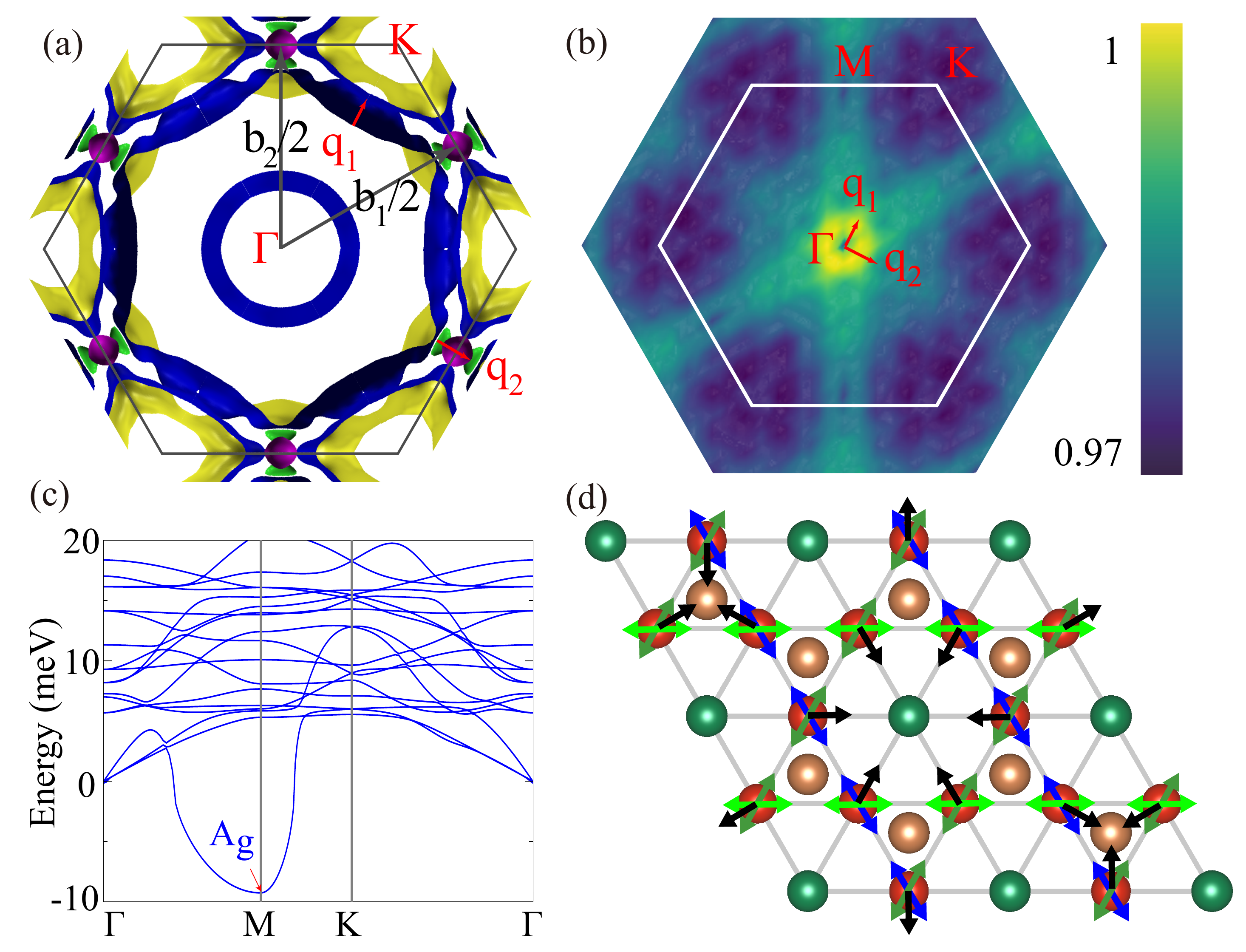}
	\caption{(a) The top view of the FSs of KV$_3$Sb$_5$ along the (001) direction.
		Two red arrows represent two small nesting vectors $\vec{q_1}$ and $\vec{q_2}$.
		(b) Normalized 2D Lindhard response function $\chi_0(q)$ in the $b_1$-$b_2$ plane.
		(c) The phonon spectrum of pristine CsV$_3$Sb$_5$.
		(d) The $A_g$ vibration modes of V atoms at three M points.
            The blue, orange and green arrows represent the vibration at $M_1$ ($\pi$, 0, 0), $M_2$ (0, $\pi$, 0) and $M_3$ ($\pi$, $\pi$, 0) points, respectively.
            The black arrows respect the total vibration direction.
	}
	\label{fig2}
\end{figure}

   According to the above calculations, we will study the structural instability of $A$V$_3$Sb$_5$ just based on the NM calculations for different supercells and structural distortions.
   First, let us focus on the structural instability of KV$_3$Sb$_5$.
   For this purpose, we construct a 2$\times$2 supercell from the pristine phase, and lower its symmetry by adjusting the coordinates of V atoms.
   As a result, seven structural configurations as shown in Fig.~1(b)-(h) are obtained.
   We notice that the T\&H phase in Fig.~1(b) and David star (DS) phase in Fig.~1(c) both satisfy the point group $D_{6h}$, and Figs.~1(d)-(h) satisfy  $C_{6h}$, $D_{3d}$, $D_{3h}$, $C_{3h}$ and $C_{3v}$ point group, respectively.
   In order to find the most stable structure, all the above superlattices and also the pristine structure of KV$_3$Sb$_5$ are optimized.
   The calculated energies are listed in TABLE \Rmnum{1}.
   It demonstrates that the T\&H phase with the shrunk trimers and hexamers of the V atoms is the most stable one, which well agrees with previous experiments and calculations~\cite{RN40,PhysRevLett.127.046401}.
   Further analysis shows that, after the optimization, the V atoms in many other structures tend to the same movements as that in T\&H phase.
   These results indicate that the trimerization and hexamerization of V atoms play important roles to stabilize the crystal structure in $A$V$_3$Sb$_5$.

    Very similar results and conclusions are obtained in the 2$\times$2 superlattice of CsV$_3$Sb$_5$.
    We select T\&H phase, $C_{6h}$, $D_{3d}$ and $C_{3v}$ type of structures as representatives and list the calculated results in TABLE \Rmnum{2}.
    Furthermore, starting from the 2$\times$2 structures, we also construct the corresponding 2$\times$2$\times$2 structural configurations by modulating the V atoms along $c$-direction.
    The calculated energies of the optimized 2$\times$2$\times$2 structures are the same as that of the corresponding 2$\times$2 structures, as shown in TABLE \Rmnum{2}.
    Therefore, we conclude that the energy change caused by the $c$-direction reconstruction is negligible, indicating that the CDW in $A$V$_3$Sb$_5$ is mainly come from the in-plane instability.

\begin{figure}
	\includegraphics[width=0.47\textwidth]{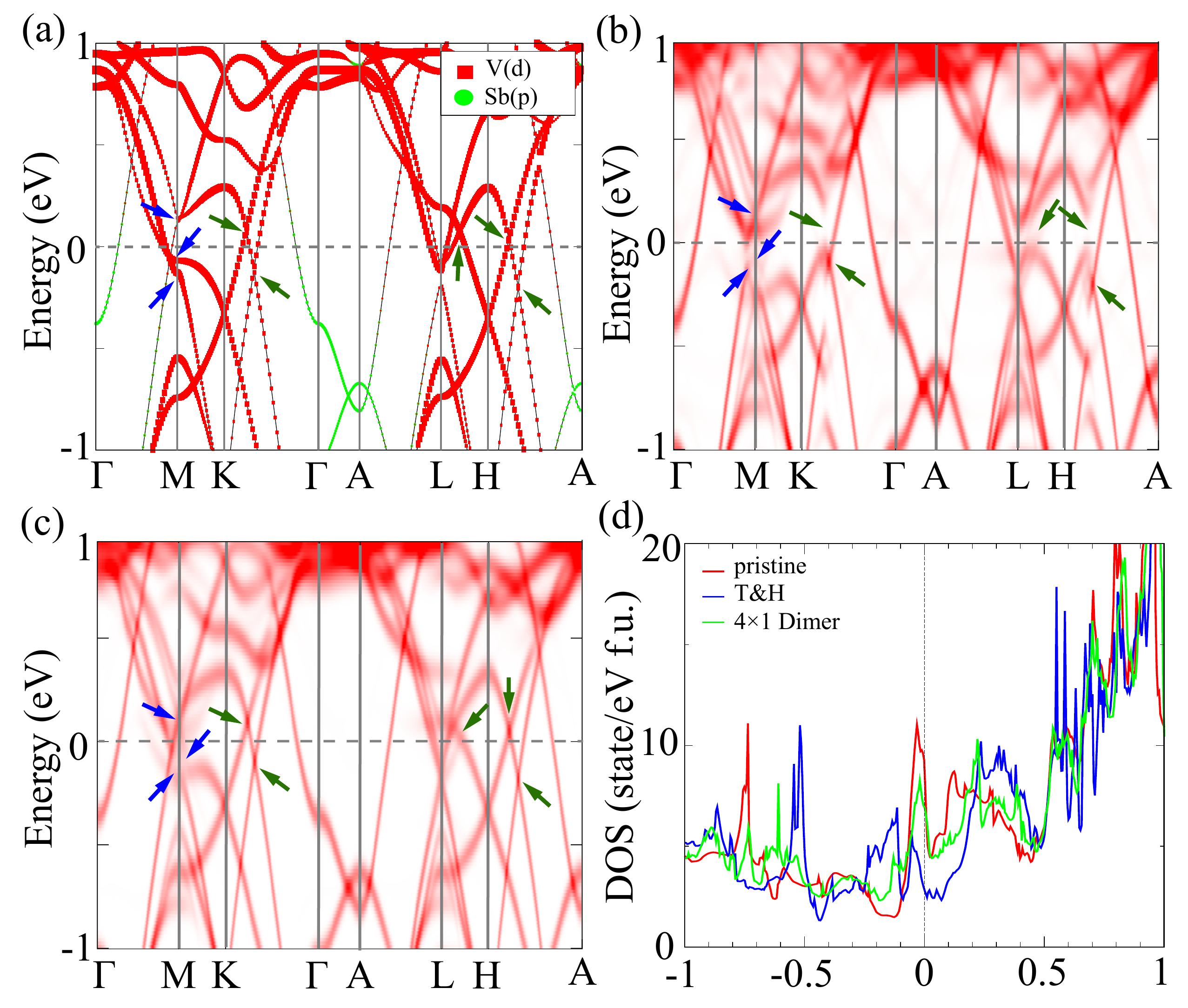}
	\caption{(a) The projected band structures of pristine CsV$_3$Sb$_5$.
		(b)-(c) The unfolded band structures of 2$\times$2 T\&H phase and 4$\times$1 Dimer phase, respectively.
		The blue and dark green arrows indicate VHS points and Dirac points, respectively.
		(d) The DOS of CsV$_3$Sb$_5$ in pristine phase (red), 2$\times$2 T\&H phase (blue) and 4$\times$1 Dimer phase (green).
	}
	\label{fig2}
\end{figure}

    Besides, a metastable 4$\times$1 supercell satisfying $C_{2h}$ point group symmetry is also discovered for the bulk and film calculations of CsV$_3$Sb$_5$, which corresponds to the experimental observations well~\cite{PhysRevB.104.075148,RN41,RN42}.
    As shown in TABLE \Rmnum{2}, the energy of such 4$\times$1 supercell is 3 meV/f.u. lower than the pristine phase and 12 meV/f.u. higher than the T\&H phase.
    More detailed analysis reveals that a dimer pattern along (010) direction, rather than the trimers and hexamers in 2$\times$2 superlattice, is formed in such 4$\times$1 supercell, as shown in Fig.~1(i).
    Moreover, a 4$\times$1$\times$2 supercell is constructed but failed to search for other kind of structural configuration.
    It always gives rise to the same energy and atomic modulation pattern as in the 4$\times$1 supercell, which further demonstrates that the in-plane instability is the main driven force of the CDW in $A$V$_3$Sb$_5$.

    In order to check whether the CDW instability originates from the FS nesting and the Peierls instability as pointed out by the recent X-ray scattering and ARPES experiments~\cite{PhysRevX.11.031050}, we calculate the FSs of the pristine KV$_3$Sb$_5$ and plot top view of them in Fig.~2(a).
    The FSs of KV$_3$Sb$_5$ have quasi-2D characteristics with a columnar FS around $\Gamma$ center, while another tri-prismatic FSs surround $K$ point.
    As a result, the FSs do not show any overlap by shifting them of vector $\vec{b_1}/2$ or $\vec{b_2}/2$, and only weak FS nesting effect can be induced by shifting two small vectors, $\vec{q_1}$ and $\vec{q_2}$, as shown in Fig.2~(a).
    Lindhard response function $\chi_0(q)$ is a straightforward evidence to reflect the FS nesting instability~\cite{PhysRevMaterials.3.024002}.
    We calculate the 2D renormalized $\chi_0(q)$ of KV$_3$Sb$_5$ and plot it in Fig.2~(b).
    Obviously, no peaks of $\chi_0(q)$ appear at the nesting vectors $\vec{b_1}/2$ or $\vec{b_2}/2$ that correspond to the 2$\times$2 supercell.
    Fig.~2(b) just exhibits some weak peaks at small $\vec{q}$, i.e., $\vec{q_1}$ and $\vec{q_2}$ in Fig.~2(a).
    These results strongly demonstrate that the CDW instability in $A$V$_3$Sb$_5$ is not from the FS nesting effect.

    Next, we focus our attention on the phonon spectrum of $A$V$_3$Sb$_5$, and display the results of CsV$_3$Sb$_5$ to study the structural instability induced by EPC.
    The calculated phonon spectrum of pristine CsV$_3$Sb$_5$ is plotted in Fig.~2(c).
    We clarify that the softening branch phonon shown at $M(\pi, 0, 0)$ point of Fig.~2(c) has $A_g$ representation,
    which corresponds to the vibration mode of V atoms along (010) direction, as illustrated by the blue arrows in Fig.~2(d).
    Furthermore, the $A_g$ mode at other two $M$ points, i.e. $(0, \pi, 0)$ and $(\pi, \pi, 0)$, give rise to the vibration of V atoms along (100) and (110) direction, respectively.
    Combining the $A_g$ modes at three $M$ points, the total vibration modes are consistent with movements of V atoms in T\&H or DS phase.
    In order to strengthen such softening mode, one natural manner is to shorten the distance between V atoms as occurred in the T$\&$H phase (see the black arrows in Fig.~2(d)).
    Thus our results demonstrate that the EPC plays crucial roles in the formation of the 2$\times$2 or 2$\times$2$\times$2 CDW phase in $A$V$_3$Sb$_5$, which is consistent with the very recent neutron scattering experiment~\cite{xie2021electronphonon} very well.
    On the other hand, the 4$\times$1 instability seems more complex.
    Electron correlation and surface instability may need to be taken into account in order to well address this problem~\cite{PhysRevB.104.075148,RN41,RN42}.

\begin{figure}
	\includegraphics[width=0.45\textwidth]{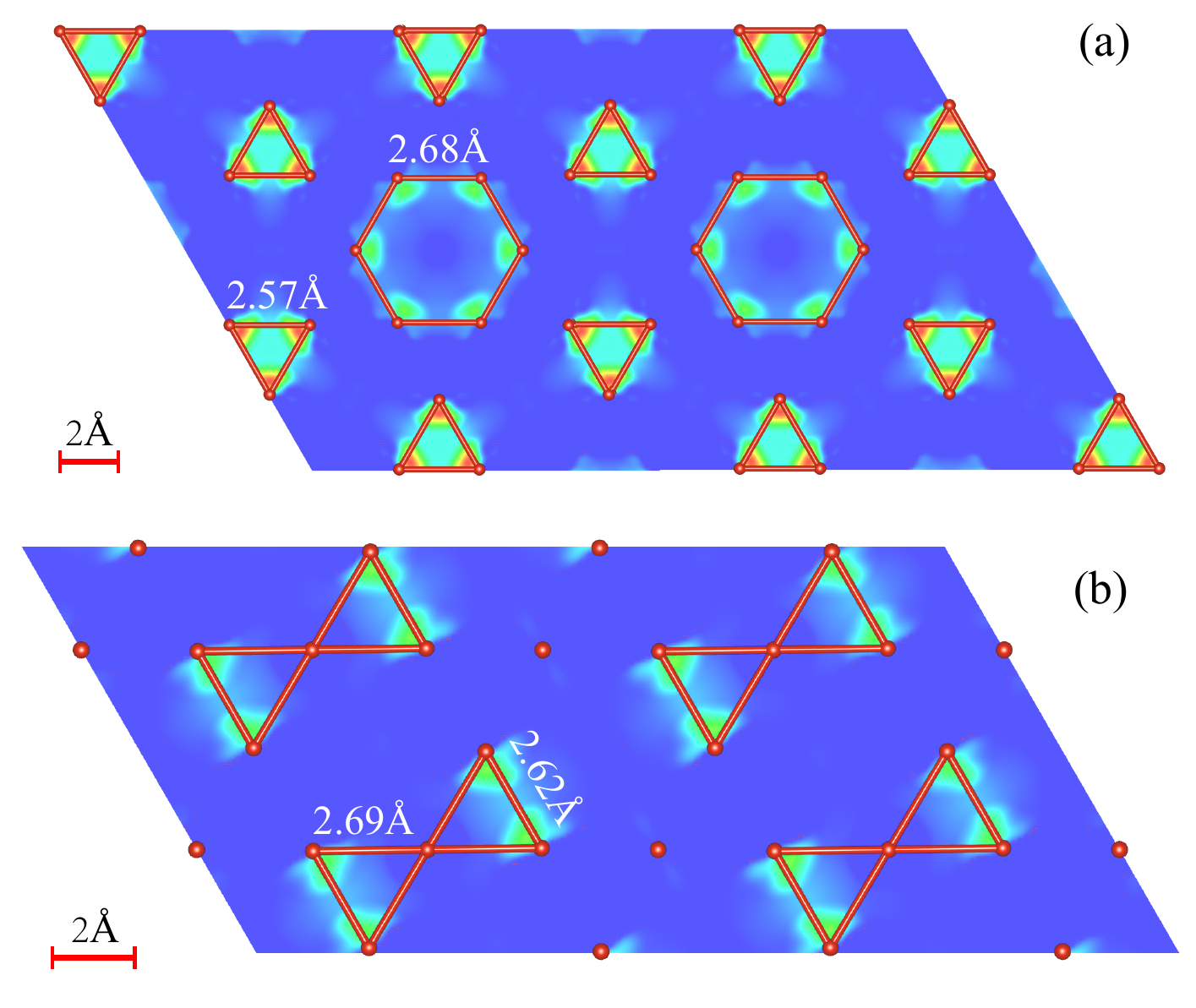}
	\caption{(a)-(b) The differential charge density distribution in V Kagome layer of CsV$_3$Sb$_5$ in 2$\times$2 T\&H phase and 4$\times$1 Dimer phase, respectively.
	}
	\label{fig4}
\end{figure}

    In the following, we would like to discuss the changes of the electronic properties between different structural phase.
    In Fig.~3(a), we plot the projected band structure of pristine CsV$_3$Sb$_5$.
    Consistent with previous reports~\cite{PhysRevMaterials.3.094407}, there are many VHS points and Dirac points close to the Fermi level, resulting in a relatively high density of states (DOS) at Fermi level ($N_{E_f}$ = 8.92 state/(eV f.u.)).
    After the 2$\times$2 reconstruction, most of the VHS points and Dirac points are gapped as shown in Fig.~3(b), which leads to a DOS transformation from Fermi level to the higher or lower energy.
    We notice that, non-zero Berry curvature is usually associated with the gapped Dirac points~\cite{RN43}, which may be the reason of the observed anomalous Hall effect~\cite{Yangeabb6003,PhysRevB.104.L041103}.
    In the metastable 4$\times$1 Dimer phase, only the spectra weight contributed by the VHS points at $M$ point is weakened, 
    while the Dirac points are marginally affected, as shown in Fig.~3(c).
    So that a weak DOS suppression is observed.
    We have summarized the change of DOS in Fig.~3(d), which demonstrates that the 2$\times$2 and 4$\times$1 reconstruction lead to different DOS suppression at Fermi level, and give rise to $N_{E_f}$ = 2.84 state/(eV f.u.) for 2$\times$2 T\&H phase and $N_{E_f}$ = 6.93 state/(eV f.u.) for 4$\times$1 Dimer phase respectively.

    Finally, we calculate the differential charge density distribution of 2$\times$2 T\&H and 4$\times$1 Dimer CsV$_3$Sb$_5$, i.e., the real space charge difference between the CDW phase and pristine phase, which reveals the real space charge modulation directly and can be compared with the STM observations qualitatively.
    The differential charge density distribution in V Kagome layer is plotted in Fig.~4.
    In the 2$\times$2 T\&H phase, the bonds of V trimers (2.57~\AA) and hexamers (2.68~\AA) are shorter than that of the pristine Kagome lattice (2.72~\AA).
    Particularly, Fig.~4(a) illustrates that the charge density in the V trimers is higher than that in the V hexamers, indicating a stronger charge assembling effect in the trimer, which agree with previous experimental and theoretical results well~\cite{FENG20211384}.
    On the other hand, the charge density modulation in 4$\times$1 Dimer phase is quite different from that in the 2$\times$2 T\&H phase.
    As illustrated in Fig.~4(b), the V atoms in the 4$\times$1 Dimer phase tend to form the V-V dimers along (010) direction with bond length 2.62~\AA.
    These results indicate that there may exist the competition between dimerization and trimerization in CsV$_3$Sb$_5$.
    More importantly, a twofold symmetric bowtie shaped charge accumulation emerges as shown in Fig.~4(b).
    Such totally new charge modulation has never been reported previously and may be responsible to the twofold resistivity anisotropy in CsV$_3$Sb$_5$~\cite{RN42}.
    Lastly, we would like to notice that, the charge modulation in 4$\times$1 phase is somewhat weaker than that in the 2$\times$2 T\&H phase, which is consistent with the analysis of their electronic structures in Fig.~3.

    In summary, our results demonstrate that the most stable structure in $A$V$_3$Sb$_5$ is the 2$\times$2 supercell with trimerization and hexamerization of V atoms.
    The phonon spectrum and Lindhard function calculations strongly suggest that such CDW phase is mainly driven by phonon instability through EPC instead of the FS nesting effect,
    which agree with many latest experimental observations very well~\cite{YuL2021,PhysRevLett.127.046401,FENG20211384,PhysRevX.11.031050}.
    In particular, a metastable 4$\times$1 phase with V-V dimer pattern is also discovered in CsV$_3$Sb$_5$,
    in which a twofold symmetric bowtie shaped charge modulation is first reported and need further experimental confirmation.

We thanks the valuable discussion with JianZhou Zhao.
This work is supported by the Ministry of Science and Technology of China (No. 2018YFA0307000)
and the National Natural Science Foundation of China (No. 11874022).
Work at Princeton University is supported by the Gordon and Betty Moore Foundation (GBMF4547 and GBMF9461).

\end{document}